\documentstyle[preprint,eqsecnum,aps]{revtex}
\begin{document}
%\tighten

\newcommand {\beq} {\begin{equation}}
\newcommand {\eeq} {\end{equation}}
\newcommand {\beqa} {\begin{eqnarray}}
\newcommand {\eeqa} {\end{eqnarray}}
\newcommand {\nextline} {\nonumber \\ & & \mbox{\hspace{.2in}} }
\newcommand {\nexteq} {\nonumber \\ & = &}
\newcommand {\half} {\frac 1 2}
\newcommand {\br}  {{\bf r}}
\newcommand {\bq}  {{\bf q}}
\newcommand {\bQ}  {{\bf Q}}
\newcommand {\corr}[2] { T\{{#1}(\br,t) , {#2}(0,0) \} }
\newcommand {\jt}  {{j}^{\rm T}}
\newcommand {\bj}  {{\bf j}}
\newcommand {\cH}  {{\cal H}}
\newcommand {\cHr}[1]  {{\cal H}_{\rm_{#1}}}
\newcommand {\eqr}[1]  {Eq.~(\ref{#1})}
\newcommand {\eqrs}[2]  {Eqs.~(\ref{#1}) and (\ref{#2})}
\newcommand {\figr}[1] {Fig.~\ref{#1}}
\newcommand {\epsq}  {\hat{\bf \epsilon}_{\bq}}
\newcommand {\zhat}  {\hat{\bf z}}
\newcommand {\qhat}  {\hat{\bf q}}

\title{
Spin Susceptibility and Gap Structure of the \\
Fractional-Statistics Gas
}
\author{J. L. Levy}

\address{
 Department of Physics, Stanford University, Stanford, CA 94305
}

\author{R. B. Laughlin}

\address{
Department of Physics, Stanford University, Stanford, CA 94305 \\
and Lawrence Livermore National Laboratory,  \\
P.O. Box 808, Livermore, California 94550
}
\maketitle
\begin{abstract}
This paper establishes and tests procedures which can
determine the electron energy gap of the high-temperature
superconductors using the $t\!-\!J$ model with spinon and holon
quasiparticles obeying fractional statistics.
A simpler problem with similar physics,
the spin susceptibility spectrum of
the spin 1/2 fractional-statistics gas,
is studied.
Interactions with the density oscillations of the system substantially
decrease the spin gap to a value of $(0.2 \pm 0.2)$
$\hbar \omega_c$, much less than the mean-field value of $\hbar\omega_c$.
The lower few Landau levels
remain visible, though broadened and shifted,
in the spin susceptibility.
As a check of the methods,
the single-particle Green's function of the
non-interacting Bose gas viewed in the fermionic representation,
as computed by the same approximation scheme,
agrees well with the exact results.
The same mechanism would reduce the gap of the $t\!-\!J$ model
without eliminating it.
\end{abstract}


\section{Introduction}  \label{secintro}

In this paper, we investigate the excitation spectrum of the spin $\half$
fractional-statistics gas, focusing upon the minimum energy
required to flip a spin.
We develop techniques enabling the computation of the energy gap
of the cuprate superconductors in terms of the $t\!-\!J$ model
possessing spinon and holon quasiparticles obeying fractional
statistics~\cite{overview}.  Using that framework,
Tikofsky, Laughlin, and Zou~\cite{tikofsky} determined
the optical conductivity and
other quantities in good agreement with exact diagonalizations of
the $t\!-\!J$ Hamiltonian for small systems.
However, their computed electron energy gap is roughly five times
larger than the experimental limit~\cite{experiment} for the cuprates.
In addition, their linear response functions displayed unreasonable
sharp structure, which had to be smoothed out in an arbitrary fashion.
The spin-one excitation spectrum of the
fractional-statistics gas provides insight into the electron gap of
the more complicated  $t\!-\!J$ model.
Specifically, an electron hole in the $t\!-\!J$ model
consists of a composite of two particles, a spinon and a holon,
each obeying fractional statistics, just as the spin-one excitation of
the fractional-statistics gas is a composite of a spin-up particle and
a spin-down hole.
As will be demonstrated later,
interactions with phonons, the density oscillations of the
fractional-statistics gas,
broaden the spin susceptibility spectrum,
thereby lowering the spin gap, the minimum spin-one
excitation energy.
As a possible comparison, Girvin and collaborators~\cite{girvin}
found the exact ground state of the spin $\half$
fractional-statistics Hamiltonian with an attractive delta-function
interaction.
However, they neglected the broadening of the spin susceptibility
spectrum and consequently obtained an overly large spin gap.

Let us define the system being studied.
The non-interacting fractional statistics gas~\cite{hannahf} with spin
corresponds to the spin $S$
fermionic Hamiltonian,
\beq {\cal H}_{\rm anyon} = \sum_{i=1}^{N} \frac{1}{2m}
 {\left| {\bf P}_i + \frac{e}{c}  {\bf A}_i  \right|}^2  \; ,
\label{anyonH}  \eeq
where the gauge field ${\bf A}_i$ has the value
\beq {\bf A}_i =  \sum_{j\neq i}^{N} {\bf A}_{ij} =
 (1\!-\!\nu) \frac{\hbar c}{e} \hat{\bf z} \times
  \sum_{j\neq i}^{N} \frac{ \br_i - \br_j }
	{ \left|  {\br_i - \br_j} \right| ^ 2 }
 \left( 1 - e^{-\alpha |\br_i - \br_j|^2} \right)
\;. \eeq
and $\br_i$ denotes the two-dimensional vector locating the
$i$th particle.
In the problem being studied, the particles have spin $S = \half$ and
statistics $\nu = \half$.
The cutoff of the gauge field $\bf A$ at small distances
prevents a weak spin instability investigated by
B\'{e}ran and Laughlin~\cite{beran}.
Setting the cutoff parameter $\alpha$ equal to $10^4$ avoids the
Hartree-Fock instability described in that paper with a negligible
effect upon the subsequent spin gap calculations.
In the lattice system to which the fractional-statistics formalism
will be applied, the gauge force would be cut off at a distance of
one lattice spacing.

The spin susceptibility,
defined for an unpolarized spin $\half$ system to be
\beqa  \chi(\br_1 \: t_1 \:|\: \br_2 \: t_2) & = &
-i < \Phi \:|\: T \{ \sigma^z(\br_1 \: t_1) \: \sigma^z(\br_2 \: t_2)
     \}  \:|\: \Phi >
\nexteq
-2i < \Phi \:|\: T \{ \sigma^-(\br_1 \: t_1) \: \sigma^+(\br_2 \: t_2)
     \}  \:|\: \Phi >   \label{susc}  \;, \eeqa
determines the spin gap of the fractional-statistics gas.
The wavefunction $|\Phi\!\!>$ is the ground state, and
the operator time dependence follows the Heisenberg picture,
\beq  \hat{\cal O}(t) =
 e^{(i/\hbar) {\cal H} t}  \hat{\cal O}  e^{-(i/\hbar) {\cal H} t}
\label{heisrep}  \;.  \eeq
The $z$-component of spin is
\beq \sigma^z(\br) =
  \Psi_{\uparrow}^{\dagger}(\br) \Psi_{\uparrow}(\br) -
  \Psi_{\downarrow}^{\dagger}(\br) \Psi_{\downarrow}(\br),
 \label{spinz} \eeq
while the spin-raising and spin-lowering operators
are defined by
\beq \sigma^+(\br) =
  \Psi_{\uparrow}^{\dagger}(\br) \Psi_{\downarrow}(\br)
 \label{spinraise} \eeq
and
\beq \sigma^-(\br) =
  \Psi_{\downarrow}^{\dagger}(\br) \Psi_{\uparrow}(\br)
  \;,  \eeq
where $\Psi_{\sigma}^{\dagger}(\br)$ and $\Psi_{\sigma}(\br)$ are
fermionic operators creating or destroying a spin $\sigma$ particle at
the position $\br$.
The Fourier transform of the susceptibility,
\beq  \chi({\bf Q},\omega) =
 L^2 \int \!\!\! \int e^{-i( {\bf Q} \cdot \br - \omega t)} \,
  \chi(\br \: t \:|\: 0 \: 0) \, dt \, d{\bf r}  \;,   \eeq
determines the spin gap $\Delta$, the smallest positive frequency $\omega$
 for which
${\rm Im} \, \chi({\bf Q},\omega)$ is non-zero at some momentum $\bf Q$.

Finding the spin gap
 at the mean-field level is straightforward.
The system may be approximated by the mean-field Hartree
Hamiltonian ${\cal H}_0$,~\cite{hannahf}
\beq {\cal H}_0 = \sum_{i=1}^{N} \frac{1}{2m}
 {\left| {\bf P}_i + \frac{e}{c}  {\bar {\bf A}}_i  \right|}^2
\label{H0}   \; ,  \eeq
corresponding to particles in a constant magnetic field
\beq  {\bar{\bf A}} = \half {\bf B} \times \br \;,
\label{abar}  \eeq
where
\beq  {\bf B} =
 (1 \!-\! \nu) \frac{hc}{e} {\bar \rho} \: \hat{\bf z} \eeq
and ${\bar \rho}$ is the average particle density of all spins.
The resulting energy spectrum consists of sharp Landau levels,
with a gap between Landau levels of the cyclotron energy,
\beq  \hbar\omega_c = \frac{\hbar eB}{mc} =
   2\pi(1 \!-\! \nu) \frac{\hbar^2}{m} \bar {\rho}
\;,  \label{omegac}  \eeq
as displayed in \figr{landaul}.
For spin $\half$ semions, the lowest Landau level is occupied with
particles of both spins, and the other levels are empty.
Thus, the spin-one excitation with the smallest energy
consists of removing
a spin-down particle from Landau level 0 and adding a spin-up particle
to Landau level 1, implying a mean-field spin gap of $\hbar\omega_c$.
For any momentum $\bQ$, the imaginary part of the spin susceptibility
 ${\rm Im} \, \chi (\bQ,\omega)$
 consists of delta functions at the frequencies $\omega$
equal to multiples of the cyclotron energy $\omega_c$.
The mean-field spin susceptibility is represented as
 the thick vertical lines in \figr{spinsusc},
in which, for $Q=0.6/a_0$, the height of each line is proportional to
the relative strength of its delta function.
For smaller $Q$, the delta function at $\omega = \omega_c$
dominates the others to an even greater extent.

Our final result for the spin
susceptibility of the fractional-statistics gas differs
substantially from the mean-field value.
\figr{spinsusc} displays the imaginary part of the spin
susceptibility, calculated in Section \ref{parthole},
as a function of the frequency $\omega$ for
various values of the momentum $Q$.
For any momentum, the lowest energy excitation has an energy
$\Delta$ of $0.32\,\hbar\omega_c$.
Actually, the calculation has rather large error bars,
as discussed in section \ref{secsec},
and we only assert that the spin gap is less than
$0.4 \,\hbar\omega_c$.
A numerical approach is needed to determine the spin gap more precisely.
In any case, the spin gap, if it exists,
is substantially less than the mean-field value
of $\hbar\omega_c$.
The susceptibility spectrum has broadened considerably, though
half of the spectral weight lies within $0.01 \, \omega_c$ of
the spin gap $\Delta$.
At larger momenta, a definite peak appears at the frequency
$\omega=1.2\, \omega_c$.

Interactions with phonons, the density oscillations of the system,
broaden the spin susceptibility spectrum into the form displayed in
\figr{spinsusc}.
Consider the effect of the phonons upon the
single-particle density of states of the
unoccupied Landau levels $n\geq 1$.
As illustrated in \figr{decayproc}(a),
a particle in Landau level 2 may emit a phonon and decay into Landau level 1,
resulting in the broadening of Landau level 2.
Also, assuming Landau level 1 is broadened, as in \figr{decayproc}(b),
a particle in that Landau level
may make an intralevel transition to a lower energy state within the
Landau level and emit a phonon, in turn broadening Landau level 1
self-consistently.
This broadening extends the particle spectrum to lower energies.
Similarly, the intralevel transitions within Landau level 0
broaden the hole spectrum, so that it reaches higher energies.
Thus, the energy gap separating the particle and hole states decreases
by an amount of order one, since the theory contains no small parameters.
The phonon interaction between the particle and the hole
lowers the spin gap further.

Conceptually, a flipped spin modifies its local environment to
diminish the overall excitation energy, the spin gap.
In other words, the lowest energy state $|\bq\!\!>$ with spin one and
momentum $\bq$ may be approximated by density waves in addition
to spin waves,
\beq |\bq\!\!> \approx \left[ \sigma_{\bq}^+ -
 \sum_{\bf k} h_{\bf k} \, \rho_{\bf k} \, \sigma^+_{\bq - \bf k} \right]
 |\Phi\!\!> \;. \eeq
Here $|\Phi\!\!>$ is the ground state, $\sigma_{\bq}^+$ is the Fourier
transform
of the spin-raising operator  $\sigma^+(\br)$ defined in \eqr{spinraise},
\beq \sigma^+_{\bq} = \int d\br \, \sigma^+(\br) \,
  e^{-i \bq \cdot \br}  \;,  \eeq
and the density operator $\rho_{\bq}$ is defined in \eqr{rhoq}.
The coefficients $h_{\bf k}$ must be determined, and we have left out
terms involving more than one phonon.
In real space, the excited state $|\bq\!\!>$ may be written,
\beq |\bq\!\!> \approx \sum_i e^{-i \bq \cdot \br_i} \sigma_i^+
 \left[ Z -
 \sum_{j\neq i} h(\br_j \!-\! \br_i) \right] \;, \eeq
where $h(\br)$ is the Fourier transform of $h_{\bf k}$,
\beq h(\br) = L^2 \int\!  \frac{{d\bf k}}{(2\pi)^2} \, h_{\bf k} \,
  e^{i {\bf k} \cdot \br} \;, \eeq
$\sigma_i^+$ is the spin-raising operator on particle $i$,
and $Z$ has the value,
\beq Z = 1 - L^2 \int\! \frac{{d\bf k}}{(2\pi)^2} \, h_{\bf k} \;. \eeq
The other particles are repelled from the flipped spin present
in a higher Landau level.
At the level of a Hartree approximation, the reduced density
near the flipped spin leads to a
reduced mean gauge field and a reduced cyclotron energy $\hbar\omega_c$,
and thus a reduced excitation energy.
In conclusion, the single-mode approximation of the spin-one
excitation made by Girvin {\it et.~al.},~\cite{girvin}
\beq |\bq_{\rm SMA}\!\!> = \sigma_{\bq}^+ \, |\Phi\!\!> \;, \eeq
describes the lowest lying spin-one excitation poorly.

Let us define certain conventions used throughout this paper.
We shall often use units in which
the cyclotron energy $\hbar\omega_c$,
defined in Eq.~(\ref{omegac}),
 and the magnetic length $a_0$,
\beq  a_0 = \sqrt{ \frac{hc/e}{2\pi B} }  =
  [2\pi (1 \!-\! \nu) \bar{\rho}]^{-1/2}   \label{a0}  \eeq
both equal one.
Let us define the density operator $\rho(\br)$,
\beq \rho(\br) = \sum_{i=1}^N \delta(\br-\br_i) \;, \label{rhor} \eeq
along with its Fourier transform $\rho_{\bq}$,
\beq \rho_{\bq} = \sum_{i=1}^N e^{-i \bq \cdot \br_i}  \;. \label{rhoq} \eeq
The mean-field current operator $\bj(\br)$,~\cite{dai}
\beq \bj(\br) = \sum_{i=1}^N \frac{1}{2m}
 \left\{ {\bf P}_i + \frac{e}{c} \bar{\bf A}_i, \delta(\br-\br_i) \right\}
\;,  \label{jr} \eeq
has a Fourier transform ${\bf j}_{\bq}$,
\beq {\bf j}_{\bq} =
 \sum_{i=1}^N \frac{1}{2m} \left\{ {\bf P}_i + \frac{e}{c} \bar{\bf A}_i ,
    e^{-i\bq\cdot\br_i}  \right\}  \;,  \label{jq} \eeq
and a transverse component,
\beq \jt_{\bq} =
{\bf j}_{\bq} \cdot (\hat{\bf z} \times \hat{\bf q})
\;, \label{jqt}  \eeq
where $\bar{\bf A}$ represents the constant magnetic field
of \eqr{abar}.

\section{Effective interaction}   \label{effintsec}

Interactions with phonons, the density oscillations of the
fractional-statistics system, broaden the spin-susceptibility
spectrum and lower the spin gap.
To analyze the system, we shall replace the
Hamiltonian $\cH_{\rm anyon}$, \eqr{anyonH},
with an effective Hamiltonian
consisting of particles in a magnetic field coupled to phonons.
The phonon system accurately describes the spin-one excitation
spectrum of the fractional-statistics gas.
Perturbative techniques will then determine the spin
susceptibility and spin gap.

The effective Hamiltonian ${\cal H}_{\rm eff}$
modelling the fractional statistics gas has the form
 \beq {\cal H}_{\rm eff} =
\sum_{i=1}^{N} \frac{1}{2m}  {\left| {\bf P}_i
+ \frac{e}{c}  {\bar {\bf A}}_i   \right|}^2   + \sum_{\bq} \half
\hbar \omega_q   \left( a_{\bq}^{\dagger} a_{\bq} + a_{\bq}
a_{\bq}^{\dagger} \right)  + \sum_{\bq}  \frac{ \alpha_q}
{\sqrt{2\omega_q}} \, \jt_{\bq}  \left( a_{\bq}^{\dagger} + a_{\bq}
\right) \label{Hph}   \;.  \eeq
The operators $a_{\bq}^{\dagger}$
and $a_{\bq}$ create and destroy a phonon of momentum $\hbar\bq$
and energy $\hbar \omega_{q}$.
The particles lie in the constant magnetic field
${\bar {\bf A}}$ given by \eqr{abar},
arising from the mean-field Hamiltonian $\cHr{0}$, \eqr{H0}.
The mean-field transverse-current
operator $ \jt_{\bq}$, defined in \eqr{jqt},
couples a particle to a phonon with coupling strenth $\alpha_{q}$.
As in the mean-field Hamiltonian $\cHr{0}$, the particle
density corresponds to filling the lowest Landau level due to the
magnetic field $\bar{\bf A}$ with particles of both spins.
The phonon dispersion $\omega_q$ and coupling $\alpha_q$, along with
the justification of the effective Hamiltonian $\cHr{eff}$, are
given below.

Let us first approximate the part of the Hamiltonian left out of the
mean-field treatment as a two-body interaction~\cite{dai},
neglecting the three-body terms which have a limited effect upon the spin
gap.
The perturbation Hamiltonian ${\cH}_1$, defined by subtracting
the mean-field Hamiltonian, Eq.~(\ref{H0}),
 from the full fractional-statistics
Hamiltonian, Eq.~(\ref{anyonH}), in the limit of large $\alpha$,
may be written,
\beqa  \cH_1 & = & \cH_{\rm anyon} - \cH_0
\nexteq
\frac{e}{c} \sum_{\bq\neq 0}
{\bf V}(\bq) \cdot {\bf j}_{-\bq} \, \rho_{\bq}
  - \frac{1}{2m} \left( \frac{e}{c} \right)^2 \sum_{\bq\neq 0}
  \sum_{{\bf p}\neq 0} {\bf V}(\bq) \cdot {\bf V}({\bf p}) \,
    \rho_{{\bf p}-\bq} \, \rho_{\bq} \, \rho_{-\!{\bf p}}   \;. \eeqa
The potential ${\bf V}(\bq)$ is defined to be,
\beq  {\bf V}(\bq) =
 (1 \!-\! \nu) \frac{1}{iL^2} \frac{hc}{eq}
   \left( {\hat{\bf z}} \times {\hat{\bq}} \right)
\;, \eeq
and the density operator $\rho_{\bq}$ and the mean-field current
operator ${\bj}_{\bq}$ are defined in
Eqs.~({\ref{rhoq}}) and (\ref{jq}).
The second term of $\cH_1$, the three-body interaction, is largest
for $\bf p$ equal to $\bq$, where the density operator
$\rho_{{\bf p}\!-\!\bq}$ reduces to the total particle number $N$.
Removing the terms with $\bf p$ not equal to $\bq$ leaves the
two-body interaction Hamiltonian,
\beqa  {\cal H}_1^{\rm 2-body} & = &
\frac{e}{c} \sum_{\bq\neq 0}
 {\bf V}(\bq) \cdot {\bf j}_{-\!\bq} \, \rho_{\bq}
  - \frac{N}{2m} \left( \frac{e}{c} \right)^2 \sum_{\bq\neq 0}
  \left| {\bf V}(\bq) \right|^2  \rho_{\bq} \, \rho_{-\!\bq}
\nexteq
\half \left( \begin{array}{ccc}
\rho_{-\!\bq} & {j}_{-\!\bq}^x & {j}_{-\!\bq}^y
\end{array} \right)
  {\cal V}^0 (\bq)
\left( \begin{array}{c}
\rho_{\bq} \\ {j}_{\bq}^x \\ {j}_{\bq}^y
\end{array} \right)  \;,  \eeqa
where, for $\bq$ in the $\hat {\bf x}$ direction,
the bare interaction ${\cal V}^0$ has the form,
\beq  {\cal V}^0(\bq)  =  \half \frac{2\pi}{L^2} \frac{1}{q^2}
%\frac{N}{m} \left( \frac{e}{c} \right)^2 |V(\bq)|^2 & 0 &
% -\frac{e}{c} V(\bq)    \\    0 & 0 & 0  \\
%\frac{e}{c} V(\bq)  & 0 & 0
 \left( \begin{array}{ccc}
1 & 0 & iq \\ 0 & 0 & 0 \\ -iq & 0 & 0
\end{array} \right)  \;.  \label{V0} \eeq
In the preceding equation and the remainder of this paper,
the cyclotron energy $\hbar \omega_c$, \eqr{omegac},
and the magnetic length $a_0$, \eqr{a0},
are both set equal to one.

Various higher order processes modify the effective interaction
between any two particles, in a fashion
similar to the screening of the Coulomb interaction in an electron gas.
Diagrammatically, the effective interaction is the sum of a bare
interaction and a polarization bubble attached to two bare interactions,
\beq  \label{VVDV}  {\cal V}(\bq, \omega) =  {\cal V}^0(\bq) +
  {\cal V}^0(\bq) {\cal D}(\bq,\omega) {\cal V}^0(\bq)  \;. \eeq
The full polarization bubble ${\cal D}$ is the correlation function
\beqa  {\cal D}(\bq, \omega)  & = &
  L^2 \int \frac{1}{i\hbar} \int_0^{\infty} dt \, d\br \,
  e^{-i (\bq \cdot \br - \omega t)} \, e^{-\eta t}
\nonumber \\ & & \mbox{\hspace {0in}} \times
{
<\! \Phi |\!\left(\!\!  \begin{array}{ccc}
\corr{\rho}{\rho} &  \corr{\rho}{j^x}  &  \corr{\rho}{j^y} \\
\corr{j^x}{\rho}  &  \corr{j^x}{j^x} &  \corr{j^x}{j^y} \\
\corr{j^y}{\rho} &  \corr{j^y}{j^x}  & \corr{j^y}{j^y}
\end{array}  \!\!\right)\!   |\Phi\!> }
\;. \label{corrfun} \nextline
  \eeqa
The density operator $\rho(\br)$ and
the mean-field current operator $\bj(\br)$ are given by
Eqs.~(\ref{rhor}) and (\ref{jr}),
the time dependence of the operators follows
the definition of the Heisenberg representation, Eq.~(\ref{heisrep}),
and the operators are time ordered.
Eq.~(\ref{VVDV}) for the effective interaction may be expressed
 in terms of the proper polarization ${\cal D}^P$,
the Feynman diagrams in the polarization
which remain connected after removing any
single interaction line, in the form,
\beq  {\cal V}(\bq, \omega) =  {\cal V}^0(\bq) +
  {\cal V}^0(\bq) {\cal D}^P(\bq,\omega) {\cal V}(\bq,\omega)
\;,  \label{VV2}  \eeq
as displayed in \figr{effint}(a).
The bare interaction ${\cal V}^0$, Eq.~(\ref{V0}), is denoted
by a thin wavy line,
the effective interaction ${\cal V}$ is denoted by the
thick wavy line,
and the proper polarization ${\cal D}^P$ is denoted by the black square.

Let us calculate the effective interaction using the Random Phase
Approximation.
In the Random Phase Approximation (RPA),
the proper polarization ${\cal D}^P$ in \eqr{VVDV}
 is estimated by the bare polarization
 ${\cal D}^0$ of the mean-field Hamiltonian $\cH_0$, \eqr{H0}.
As displayed diagrammatically in \figr{effint}(b),
\beq  {\cal V}^{\rm RPA}(\bq, \omega) =  {\cal V}^0(\bq) +
  {\cal V}^0(\bq) {\cal D}^0(\bq,\omega) {\cal V}^{\rm RPA}(\bq,\omega)
\label{vrpa}  \;,  \eeq
where the straight lines represent single-particle propagators
of the Hamiltonian $\cH_0$.
The bare polarization ${\cal D}^0$ is found by
replacing $|\Phi\!\!>$ with $|\Phi_0\!\!>$, the ground state of $\cH_0$,
in Eq.~(\ref{corrfun}) and using the Hamiltonian $\cH_0$ in the
Heisenberg representation, Eq.~(\ref{heisrep}).
The calculation of the bare polarization of a spinless system in
\onlinecite{dai} and \onlinecite{fetterrpa}
is modified for spin $\half$ semions to have the value,
\beq  {\cal D}^0(\bq,\omega)
= 2 \frac{L^2}{2\pi}  \left( \begin{array}{ccc}
q^2\Sigma_0     &  q\omega\Sigma_0   &  -iq\Sigma_1       \\
q\omega\Sigma_0 & \omega^2\Sigma_0\!-\!1 & -i\omega\Sigma_1   \\
iq\Sigma_1 	& i\omega\Sigma_1    & \Sigma_2
\end{array} \right)  \;,   \label{D0}  \eeq
where
\beq \Sigma_j(\bq,\omega) \equiv \sum_{n=1}^{\infty}
 \frac{e^{-b}b^{n-1}} {(n\!-\!1)! [\omega^2-(n\!-\!i\eta)^2]} (n\!-\!b)^j
  \eeq
and $b$ represents $\half q^2$.
The factor of two in Eq.~(\ref{D0}) arises from the spin
degrees of freedom.
Substituting Eqs.~(\ref{V0}) and (\ref{D0}) into Eq.~(\ref{vrpa})
results in the RPA interaction,
\beq {\cal V}^{\rm RPA}(\bq,\omega) = \half \frac{2\pi}{L^2q^2D}
\left( \begin{array}{ccc}
1+\Sigma_2 & 0 & iq(1+\Sigma_1) \\
0 & 0 & 0 \\
-iq(1+\Sigma_1) & 0 & q^2\Sigma_0
\end{array}  \right)  \;,  \label{vrpaf} \eeq
where
\beq  D = (1+\Sigma_1)^2 - \Sigma_0 (1+\Sigma_2) \;.  \eeq

For fixed $\bq$, the RPA interaction ${\cal V}^{\rm RPA}(\bq,\omega)$
contains poles at discrete values of $\omega$, with a spacing of
approximately one between poles.
\figr{imvrpa}\ graphs the imaginary part of the
current-current
 component of ${\cal V}^{\rm RPA}$ for
three values of $q$.
The solid line in each plot has been broadened with a small
value of $\eta$ for visibility.
However, that discrete Landau level structure is not physical, and would
disappear in a more accurate calculation of the proper polarization
${\cal D}^P$.
The effective interaction at finite $q$ will be broadened
over a range of energies, resembling the dashed line in
Figs.~\ref{imvrpa}(b) and \ref{imvrpa}(c).

Nevertheless, we will approximate the dynamic effective interaction
as a sharp mode
for each momentum $\bq$ in the current-current channel,
disregarding both the broadening of the effective
interaction and
the effect of the density-density and density-current channels.
For $q<2$, the lowest energy singularity
in the current-current
RPA interaction dominates the other singularities
in the three interaction channels,
while for $q>2$ the dynamic interaction is too weak to have a
significant effect.
The effective interaction then has the form,
\beq  {\cal V}^{\rm eff}(\bq,\omega)
 = {\cal V}^0(\bq) + \left( \begin{array}{ccc}
0 & 0 & 0 \\ 0 & 0 & 0 \\ 0 & 0 &
\frac {|\alpha_q|^2} {\omega^2-(\omega_q-i\eta)^2}
\end{array}  \right)  \;.  \label{veff}  \eeq
The excitation, a phonon, has an energy $\hbar\omega_q$ and
a transverse-current coupling $\alpha_q$.

Let us determine the coefficients $\alpha_q$ and $\omega_q$
appearing in \eqr{veff} for the effective interaction ${\cal V}^{\rm eff}$.
We will define a quantity $A_q$ equal to the area under the curves
in \figr{imvrpa},
\beq  A_q  \equiv \int_0^{\infty}  \left[
 -\frac{1}{\pi} \: {\rm Im} \, {\cal V}^{\rm RPA}_{JJ} (q,\omega)
\right]  d\omega  \;.  \eeq
Then $\omega_q$ is defined as
the weighted average value of $\omega$,
\beq  \omega_q \equiv \frac{1}{A_q}
  \int_0^{\infty}  \omega \left[
 -\frac{1}{\pi} \: {\rm Im} \, {\cal V}^{\rm RPA}_{JJ} (q,\omega)
\right]  d\omega  \;,  \label{omegaave}  \eeq
and the coupling strength $\alpha_q$ obeys
\beq |\alpha_q|^2 \equiv 2 \, A_q \, \omega_q  \;.  \label{alphave}  \eeq
Substituting Eq.~(\ref{V0}) into Eq.~(\ref{VV2}) demonstrates that
\beq  {\cal V}_{JJ} = \left( \frac{\pi}{L^2} \right)^2 \frac{1}{q^2}
  {\cal D}_{\rho\rho}  \;,  \eeq
which, along with the f-sum rule~\cite{pines}, reduces Eqs.~(\ref{omegaave})
and (\ref{alphave}) to
\beq  |\alpha_q|^2 = \frac{\pi}{L^2}  \;.  \label{alphaqf}  \eeq
Numerical calculations confirm this constant value of $|\alpha_q|^2$,
and fit $\omega_q$ to the formula,
 \beq  \omega_q  ^{\rm (formula)} =
   \left[ (v_sq)^4 + (q^2/2)^4 \right] ^ {1/4}
  \;,  \label{omegaqf}   \eeq
with a sound velocity $v_s$ of 1.
Both $\omega_q$ and $\left. \omega_q \right. ^{\rm (formula)}$
 are graphed in \figr{omegaqfig}.

Actually, the Random Phase Approximation does not determine the speed
of sound $v_s$ sufficiently accurately.
Together, Hartree-Fock and correlation effects~\cite{hannahf,dai,hannacorr}
lower the sound
speed of the spinless fractional-statistics gas by approximately 9\%.
The sound speed of the spin $\half$ fractional-statistics gas would
presumably shift by a comparable amount.
Inserting $v_s=0.9$ into \eqr{omegaqf} has a sizable effect upon
the resulting spin gap, which we included in the error bars of the
spin gap calculation in Section \ref{secsec}.

The static interaction ${\cal V}^0$ in the effective interaction
${\cal V}^{\rm eff}$, \eqr{veff}, does not affect the spin gap.
When calculating the spin susceptibility diagrammatically, as in
\figr{kohn},
each of the phonon interaction lines is short-ranged.
Dampening the interaction
${\cal H}_{\rm anyon} \!-\! {\cal H}_0$
 for large particle separations
does not affect those Feynman diagrams.
By Kohn's theorem~\cite{kohn},
the remaining short-ranged interactions between particles
in a magnetic field do not change the excitation energy at small momenta
from the value without an interaction,
the cyclotron energy $\hbar \omega_c$.
Thus, the static interaction ${\cal V}^0$
does not change the $Q\!\rightarrow\! 0$ spin gap
and may be omitted.
B\'{e}ran and Laughlin~\cite{beran} demonstrated that with a cutoff
parameter $\alpha$ in the fractional-statistics Hamiltonian
$\cHr{anyon}$, \eqr{anyonH}, corresponding to a reasonable doping of
a lattice model, the static interaction ${\cal V}^0$ does not create
a roton to lower the spin gap at finite~$Q$.
Thus, the fractional-statistics system may be described as the mean-field
Hamiltonian~$\cHr{0}$, \eqr{H0},
and a dynamic transverse-current interaction, the second term in
\eqr{veff},
all of which reduces to the effective Hamiltonian $\cHr{eff}$, \eqr{Hph}.

In summary, we have obtained a simpler system $\cHr{eff}$, \eqr{Hph},
which possesses the same
spin gap as the fractional-statistics gas $\cHr{anyon}$, \eqr{anyonH}.
The effective Hamiltonian consists of particles in a magnetic field
coupled transversely to phonons.
The phonon dispersion $\omega_q$ is given by \eqr{omegaqf}, and
the particle-phonon coupling $\alpha_{q}$ is given by \eqr{alphaqf}.
In the remainder of this paper, we will investgate the effective
Hamiltonian $\cHr{eff}$ to find the spin gap.

\section{Single-particle density of states}   \label{singpart}

Let us compute the single-particle density of states of the
fractional-statistics gas.
The density of states allows a crude estimate of the spin
gap, since interactions between a particle and a hole present in
a spin-one excitation lower the spin gap further.
However, the single-particle computation clearly illustrates the
mechanisms of broadening and gap lowering present in the spin
susceptibility spectrum.
Also, the formal methods used to compute the spin
susceptibility in section \ref{parthole} resemble the techniques
used to compute the density of states.
We will start with the effective Hamiltonian $\cHr{eff}$, \eqr{Hph},
consisting of particles in a magnetic field coupled to phonons.
As displayed in \figr{decayproc}, a particle in any Landau level may
decay into either a lower Landau level or a lower energy state in the
same Landau level, emitting a phonon.
In this section, we will find the extent of the spectrum broadening
induced by that decay process.

The density of states depends upon the single-particle Green's
function of the fractional statistics gas,
\beq G(\br_1 t_1 \sigma_1 | \br_2 t_2 \sigma_2) =
  -i < \Phi \:|\: T \{ \Psi_{\sigma_1}(\br_1,t_1) \:
	\Psi_{\sigma_2}^{\dagger}(\br_2,t_2)
     \}  \:|\: \Phi >   \;, \label{greenm}  \eeq
where $|\Phi\!\!>$ is the ground state. The operators
$\Psi_{\sigma}^{\dagger}(\br,t)$ and
$\Psi_{\sigma}(\br,t)$ create and destroy a spin $\sigma$ particle
at position $\br$ and time $t$ in the Heisenberg representation,
\eqr{heisrep}.
The particle may emit and later absorb a virtual phonon, giving the
Green's function a complex self-energy which broadens the spectrum
and lowers the spin gap.
As illustrated in \figr{dysons}, we will solve Dyson's equation for the
exchange of a single phonon in
the effective Hamiltonian $\cHr{eff}$, \eqr{Hph}, to find the spin gap.

Let us first find the Green's function corresponding to
 the bare Hamiltonian $\cH_0$, \eqr{H0}, in the absence of phonons.
A constant magnetic field in the symmetric gauge
possesses the Landau level orbitals $\varphi_{nk}$~\cite{hannahf},
\beq  \varphi_{nk}(z)  = \frac{1}{\sqrt{2\pi2^{n+k}n!k!}}
  \left( \half z - 2 \frac{\partial}{\partial z^*} \right) ^n
  \left( \half z^* - 2 \frac{\partial}{\partial z} \right) ^k
  e^{-|z|^2/4}   \;, \label{llstates}  \eeq
with Landau level energies,
\beq \epsilon_{nk}^{(0)} = \epsilon_{n}^{(0)} =
 n + \half  \label{en0} \;.  \eeq
The complex number $z=x+iy$ refers to the position,
and the natural units $\hbar \omega_c$, \eqr{omegac}, and
$a_0$, \eqr{a0}, are set equal to one.
As drawn in \figr{landaul},
Landau level 0 is occupied with particles of
both spins, while the higher Landau levels are empty.
The bare Green's function may be written
in the Landau level basis as,
\beq G^0(\br_1 t_1 \sigma_1 | \br_2 t_2 \sigma_2) =
 \delta_{\sigma_1\sigma_2} \frac{1}{2\pi} \int \sum_{nk} \sum_{n'k'}
 <\! n'\:k' \:|\: G^0(E) \:|\: n\:k \!>
 \varphi_{n'k'}(\br_1)
 \varphi_{nk}^*(\br_2)  e^{iE(t_2-t_1)} dE  \;, \eeq
where
\beq <\! n' \: k' \:|\: G^0(E) \:|\: n \: k \!> =
 \delta_{nn'} \delta_{kk'} \frac{1}{E-(n+1/2)+i\eta_n} \;, \eeq
and the infinitesimal $\eta_n$ is positive for levels $n\geq 1$
and negative for level $n=0$.

The Green's function of the effective Hamiltonian $\cHr{eff}$, \eqr{Hph},
may be found by solving Dyson's equation for the exchange of a single
phonon.
This process is illustrated in \figr{dysons},
 where the thin straight line is
the bare Green's function $G^0$, the thick straight line is the full
Green's function $G$, and the thick wavy line is the phonon propagator.
We included the diagrams in which, besides the phonons,
 only a single particle or a single hole
is present at any point in time.
For instance, the exchange
graph in \figr{dysons}\ where $n\geq 1$ and $m=0$ has been neglected.
This approximation is similar to the Tamm-Dancoff
approximation~\cite{tammdancoff} in nuclear physics.
We have also left out diagrams with crossed phonon interaction lines.
The relevant particle quantum states $|n\:k\!\!>$,
in addition to the phonons, consist of Landau level 0
filled with particles of both spins, with an extra particle in the orbital
$\varphi_{nk}$ for $n\geq 1$, or a hole in the orbital $\varphi_{0k}$ for
$n=0$.

Dyson's equation involves the matrix elements of
the transverse-current operator $\jt_{\bq}$, \eqr{jqt},
between the Landau level states $|n \: k\!\!>$.
Summing over the intermediate states $|m \: p\!\!>$ within a
Landau level $m$ and
averaging over the direction $\theta$ of the phonon momentum $\bq$
%where
%\beq \bq = q\cos\theta\,{\bf \hat{x}} + q\sin\theta\,{\bf\hat{y}}
%\;, \eeq
combines those matrix elements into a simple expression.
After performing those operations, we find,
\beq -\frac{1}{2\pi} \int_0^{2\pi}  \sum_p
<\! n' \: k' \:|\: \jt_{\bq} \:|\: m \: p \!>
<\! m \: p \:|\: \jt_{-\bq} \:|\: n \: k \!> d\theta
 = \delta_{nn'} \delta_{kk'} |{\cal M}_{nm}(q)|^2  \;, \label{calm} \eeq
The effective matrix element ${\cal M}_{nm}(q)$,
which is independent of the index $k$,
has the value
\beq  {\cal M}_{nm}(q)  =  \sqrt{ \frac{m!}{2n!}} b^{(n-m-1)/2}
 \left[ n L_m^{n-m-1}(b) - b L_m^{n-m+1}(b) \right] e^{-b/2}
\label{gammal}  \;, \eeq
where $L$ denotes an associated Laguerre polynomial,
\beq L_n^{\alpha}(b) =
  \frac{1}{n!} e^b b^{-\alpha} \frac{d^n}{db^n}
 \left( e^{-b} b^{n+\alpha} \right) \;, \label{laguerre} \eeq
and $b$ stands for $\half q^2$.
%The magnitude of ${\cal M}_{nm}(q)$ is symmetric upon interchange of
%$n$ and $m$,
%\beq |{\cal M}_{nm}(q)|^2 = |{\cal M}_{mn}(q)|^2 \;. \eeq

We may now write down Dyson's equation for the single-particle
propagator arising from the exchange of a single phonon.
The Green's function will be computed in the Landau level basis,
\beq G(\br_1 t_1 \sigma_1 | \br_2 t_2 \sigma_2) =
 \delta_{\sigma_1\sigma_2} \frac{1}{2\pi} \int \sum_{nk} \sum_{n'k'}
 <\! n'k' \:|\: G(E) \:|\: nk \!>
 \varphi_{n'k'}(\br_1)
 \varphi_{nk}^*(\br_2)  e^{iE(t_2-t_1)} dE  \;. \label{gllbasis} \eeq
Because of \eqr{calm},
the full Green's function ${G}(E)$ is diagonal
in the basis of Landau level orbitals,
\beq  <\!n' \: k' \:|\: {G}(E) \:|\: n \: k \!> =
  \delta_{nn'}\delta_{kk'}  G_n(E) \;.  \label{gdiag}  \eeq
The Landau level Green's function $G_n$ has the standard form,
\beq  G_n(E) =  \frac{1}{E - (n\!+\!1/2) -\Sigma_n(E) + i\eta_n}
\label{Gneq} \;,  \eeq
with a self-energy,
\beq  \Sigma_n(E) = i L^2 \int_{-\infty}^{\infty} \frac{d\omega}{2\pi}
 \int \frac{d\bq}{(2\pi)^2}
 \sum_m \frac{|{\cal M}_{nm}(q)|^2 |\alpha_q|^2}
 {\omega ^2 - (\omega_q- i\eta)^2}  G_m(E+\omega)
 \label{selfeq} \;.  \eeq
The coupling $|\alpha_q|^2$ is given by \eqr{alphaqf}, and
the phonon energy $\omega_q$ is approximated by \eqr{omegaqf}.
When $n$ is an unoccupied Landau level, $n\geq 1$,
the index $m$ is summed over the
levels $m\geq 1$, and when $n=0$, only $m=0$ is present in the sum.

The analytic structure of the Landau level Green's function $G_m(E)$
allows a simplification of the self-energy expression, \eqr{selfeq}.
When Landau level $m$ is unoccupied, each of the poles of $G_m(E+\omega)$
lies in the lower half of the complex plane of $\omega$,
while the phonon propagator has a pole on each side of the real axis,
as indicated in \figr{contour}.
The $\omega$ integral may be evaluated by closing the contour
upwards, so that for $n\geq 1$,
\beq  \Sigma_n(E) = L^2
 \int \frac{d\bq}{(2\pi)^2}
 \sum_{m=1}^{\infty} |{\cal M}_{nm}(q)|^2
  \frac {|\alpha_q|^2} {2 \omega_q}  G_m(E-\omega_q)
\;.  \label{llself} \eeq
Landau level 0 has a similar self-energy
expression,
\beq  \Sigma_0(E) = L^2
 \int \frac{d\bq}{(2\pi)^2}
 |{\cal M}_{00}(q)|^2
  \frac{  |\alpha_q|^2} {2\omega_q}  G_0(E+\omega_q)
\;. \label{llself0} \eeq

Even though the resulting density of states is physically incorrect,
it provides a reasonable estimate of the spin gap.
We solved Eqs.~(\ref{Gneq}), (\ref{llself}), and (\ref{llself0})
 for $G_n(E)$.
In \figr{imge}, we display the density of states $D_n(E)$ within
the lowest four Landau levels,
\beq D_n(E) = \bar{\rho} \, \left| \frac{1}{\pi} \, {\rm Im} \, G_n(E)
\right| \;.  \label{dos} \eeq
Our calculation left out the effect of the static interaction,
 ${\cal V}_0$ in \eqr{veff}, which changes the density of states greatly.
The static interaction causes a gap between the occupied and
unoccupied states which diverges logarithmically with the size of
the sample.
However, due to Kohn's theorem, as argued in Section \ref{effintsec},
the static interaction does not affect the spin gap, the quantity
of physical interest.
Thus, except for leaving out the particle-hole and multiple-phonon
interactions,
the resulting spin gap remains valid.

The resulting broadening of the spectrum significantly reduces the gap energy
relative to the mean-field gap displayed in \figr{landaul}.
The broadened unoccupied states in \figr{imge} extend down to
a minimum particle energy of
\beq E_{\rm min, p}
 = \left(\epsilon_1^{(0)} - 0.48 \right) \hbar \omega_c
= 1.02 \, \hbar \omega_c  \;, \eeq
and the occupied states reach a maximum hole energy of
\beq E_{\rm max, h}
 = \left(\epsilon_0^{(0)} + 0.05 \right) \hbar \omega_c
 = 0.55 \, \hbar \omega_c\;. \eeq
If the particle-hole interactions are neglected,
the lowest lying spin-one state then has an energy,
\beq \Delta^{\rm no\ p-h} = E_{\rm min, p} - E_{\rm max, h}
 = 0.47 \, \hbar \omega_c \;. \eeq
Both the dynamic interaction between the particle and the hole
and multiple phonon interactions will lower the spin gap further.

Finally, let us examine the overall distribution of the single-particle
spectrum.
The density of states of Landau level~1 diverges as the energy
approaches $E_{\rm min, p}$ from above, so that roughly 60\% of that
level's states are within $0.01\,\hbar\omega_c$ of $E_{\rm min, p}$,
while the tail extending
to higher energies contains the rest of the states.
Similarly, 90\% of Landau level~0 lies very close to
$E_{\rm max, h}$,
with the tail containing the remaining spectral weight extending to lower
energies.
Landau level 2 contains a strong peak at an energy somewhat below
$\epsilon_2^{(0)}$, in addition to some broadened structure at both higher
and lower energies.
Landau level 3 is more broadened than the lower Landau levels,
though a peak remains visible.
Despite the broadening, the average energy of each Landau level,
\beq <\!E_n\!> \equiv \frac{1}{\bar{\rho}} \int_{-\infty}^{\infty}
 E \, D_n(E) \, dE \;,  \eeq
maintains the unperturbed value $\epsilon_n^{(0)}$.

\section{Spin susceptibility and gap}  \label{parthole}
Let us compute the spin susceptibility of the fractional-statistics
gas, and consequently its gap.
We shall find the retarded spin-one particle-hole propagator,
\beq {\cal F}(\br_1 \br_4 t_1 \:|\: \br_2 \br_3 t_2)
 = -i \theta(t_1 \!-\! t_2)
 <\! \Phi \:|\: \Psi^{\dagger}_{\downarrow}(\br_4,t_1) \:
	\Psi_{\uparrow}(\br_1,t_1) \Psi^{\dagger}_{\uparrow}(\br_2,t_2)
   \Psi_{\downarrow}(\br_3,t_2)
       \:|\: \Phi\! >  \label{Fr} .  \eeq
The particle-hole pair produced in the intermediate state may emit and
then subsequently absorb a phonon.
The phonons arise in the effective Hamiltonian $\cHr{eff}$,
\eqr{Hph}, which as argued in Section \ref{effintsec} has the same
spin gap as the fractional-statistics gas.
The single-particle self-energy, $\Sigma_n(E)$ in \eqr{llself},
accounts for the case when the particle both emits and absorbs a
phonon, and the single-hole self-energy, $\Sigma_0(E)$ in \eqr{llself0},
accounts for the case when the hole both emits and absorbs a phonon.
However, the full particle-hole propagator $\cal F$ is needed to include
the sizable interaction between the particle and the hole,
where one emits and the other absorbs a phonon.
Consequently, the spin one spectrum exhibits more broadening and
gap lowering than does a convolution of the single-particle Green's
functions computed earlier.
The spin susceptibility and spin gap are readily expressed in terms
of the pair propagator.

As in the section \ref{singpart}, we have only included
processes in which a single particle-hole pair propagates forwards
in time, emitting and absorbing phonons.
The Random Phase Approximation (RPA) Feynman graphs,
which for the spin 0 particle-hole propagator exhibit the corrections
to the Hartree-Fock ground state most strongly,
as a reversed time-ordering~\cite{dai},
do not appear in the spin 1 propagator.
Thus, the Tamm-Dancoff approximation~\cite{tammdancoff} should be reliable.
Graphs containing crossed phonon interaction lines have again been
excluded.

A magnetoexciton basis~\cite{hannaph}, describing
a particle and a hole in a magnetic field
in terms of eigenstates of both momentum and the bare Hamiltonian $\cHr{0}$,
 proves to be convenient in studying the particle-hole propagator.
The magnetoexciton particle-hole wavefunction
has the form,
\beq  \psi_{n\alpha}(z_1,z_2)  =
\frac{(-1)^n}{L\sqrt{2 \pi 2^{n} n! }}
  \left( 2 \frac{\partial}{\partial z_1^*} - \half z_1 \right) ^n
  e^{ -[|z_1|^2 + |z_2|^2 + |z_{\alpha}|^2]/4}
  e^{ [z_1^*z_2 + z_1^*z_{\alpha} - z_2z_{\alpha}^* ]/2 }
\label{magneto}  \;, \eeq
where $z=x+iy$ is a position written as a
complex number.
The spin-up particle lies in Landau level $n\geq 1$,
the spin-down hole lies in Landau level $0$,
and the pair has momentum $\bQ$, represented as a complex number
 $z_{\alpha} = iQ_x \!-\! Q_y$.
The corresponding quantum state $|n \: \bQ\!>$ is an energy eigenstate
of the mean-field Hamiltonian $\cHr{0}$, \eqr{H0}, with energy
\beq \epsilon_{n\bQ}^{(0)} = n \;. \eeq
The bare particle-hole Green's function ${\cal F}^0$
of the Hamiltonian
$\cHr{0}$ may be expressed in the magnetoexciton basis,
\beqa {\cal F}^0(\br_1 \br_4 t_1 | \br_2 \br_3 t_2) &=&
\frac{1}{2\pi} \int \sum_{n\,\bQ} \sum_{n'\,\bQ'}
 <\! n'\:\bQ' \:|\: {\cal F}^0(E) \:|\: n\:\bQ \!>
\nextline \mbox{\hspace{.5in}} \times
\psi_{n'\alpha'}(\br_1,\br_4) \, \psi_{n\alpha}^*(\br_2,\br_3) \,
e^{iE(t_2-t_1)} \, dE \;, \label{F0} \eeqa
where
\beq <\!n' \: \bQ' \:|\: {\cal F}^0(E) \:|\: n \: \bQ\!> =
 \delta_{nn'} \, \delta_{\bQ\bQ'} \, \frac {1}{E-n+i\eta} \;. \eeq

The phonon coupling appearing in the particle-hole propagator depends upon the
transverse-current matrix element between two magnetoexcitons.
The transverse-current operator $\jt_{\bq}$, defined in \eqr{jqt},
may act on either the particle or the hole,
and both couplings must be included.
The coupling between the particle and a phonon has a matrix element,
\beqa  <\! m \: \bQ\!-\!\bq \:|\: j_{\bq}^{\rm T \: (P)} \:|\: n \: \bQ\!>
 & = &
 \int\!\!\!\int d\br_1 \, d\br_2 \lim_{\br_1' \rightarrow \br_1}
  e^{-i \bq \cdot \br_1}
\nextline \mbox{\hspace{-1.7 in}} \times
 \left[ \half ({\bf P}_1 \!-\! {\bf P}_{1'}) + \bar{\bf A}(\br_1) \right]
 \cdot \epsq \:
 \psi_{m \beta}^*(\br_1',\br_2) \,
  \psi_{n \alpha}(\br_1,\br_2)  \;, \eeqa
while the coupling between the hole and a phonon has a matrix element,
\beqa  <\! m \: \bQ\!-\!\bq \:|\: j_{\bq}^{\rm T \: (H)} \:|\: n \: \bQ\!>
 & = &
 -\int\!\!\!\int d\br_1 \, d\br_2 \lim_{\br_2' \rightarrow \br_2}
  e^{-i \bq \cdot \br_2}
\nextline \mbox{\hspace{-1.7 in}} \times
 \left[ \half ({\bf P}_2 \!-\! {\bf P}_{2'}) - \bar{\bf A}(\br_2) \right]
 \cdot \epsq \:
 \psi_{m \beta}^*(\br_1,\br_2) \,
  \psi_{n \alpha}(\br_1,\br_2')
  \;. \label{jqth}  \eeqa
%The symbols 1 and 2 refer to the position variables,
The mean-field $\bar{\bf A}(\br)$ is specified by \eqr{abar},
$\epsq$ is the transverse direction
$\zhat \times \qhat$,
and $z_{\beta} = i(Q_x\!-\!q_x) \!-\! (Q_y\!-\!q_y)$.
Since the hole has the opposite charge, its interaction with both the
phonons and the magnetic field has the opposite sign.
The resulting matrix elements have the values,
\beqa  <\! m \: \bQ\!-\!\bq \:|\: j_{\bq}^{\rm T \: (P)} \:|\: n \: \bQ\!>
 & = &
 e^{-(|z_{\alpha}|^2 + |z_{\beta}|^2)/4}
 e^{z_{\alpha}^* z_{\beta} /2}
\nextline \mbox{\hspace{-1.7 in}} \times
 \frac{i}{|z_{\gamma}|}
 \sqrt{ \frac{m!}{n!}} \left( -\frac{z_{\gamma}}{\sqrt{2}} \right) ^{n-m}
 \left[n\, L_m^{n-m-1}(b) - b\, L_m^{n-m+1}(b) \right] \;,
\eeqa
and
\beq  <\! m \: \bQ\!-\!\bq \:|\: j_{\bq}^{\rm T \: (H)} \:|\: n \: \bQ\!>
  =
\delta_{mn}
 e^{-(|z_{\alpha}|^2 + |z_{\beta}|^2)/4}
 e^{z_{\alpha} z_{\beta}^* /2}
 \frac{i}{|z_{\gamma}|}
 \left[ -b \right] \;,
\eeq
where $z_{\gamma} = iq_x \!-\! q_y$, $b$ denotes $\half q^2$,
and the Laguerre polynomials $L$ are defined in \eqr{laguerre}.
The transverse-current matrix element is the sum of the particle
and the hole components,
\beqa
  <\! m \: \bQ\!-\!\bq \:|\: j_{\bq}^{\rm T} \:|\: n \: \bQ\!> &=&
  <\! m \: \bQ\!-\!\bq \:|\: j_{\bq}^{\rm T \: (P)} \:|\: n \: \bQ\!> +
  <\! m \: \bQ\!-\!\bq \:|\: j_{\bq}^{\rm T \: (H)} \:|\: n \: \bQ\!>
\nexteq
\sqrt{\frac{m!}{n!}} \frac{i}{q}
\left( \frac{-z_{\gamma}}{\sqrt{2}} \right) ^{n-m} e^{-b/2}
\nextline \mbox{\hspace{0in}}\times
\left\{ e^{(i/2) (\bq \times \bQ) \cdot \hat{\bf z}}
  \left[ n \, L_m^{n-m-1}(b) - b \, L_m^{n-m+1}(b)
 \right] \right.
\nextline \mbox{\hspace{.2in}} \left.
 - \: \delta_{nm} e^{-(i/2) (\bq \times \bQ) \cdot \hat{\bf z}}
 \: b \right\} \;. \label{ephase} \eeqa
Since small momenta dominate the interaction, we may neglect
the phases proportional to $\bq \times \bQ$ in \eqr{ephase},
approximating the current matrix element
 $ <\!\! m \: \bQ\!-\!\bq \:|\: j_{\bq}^{\rm T} \:|\: n \: \bQ\!\!>$
 by a term
${\cal M}_{nm}'(\bq)$ independent of $\bQ$,
\beq {\cal M}_{nm}'(\bq) =
\sqrt{\frac{m!}{n!}} \frac{i}{q}
\left( \frac{-z_{\gamma}}{\sqrt{2}} \right) ^{n-m}
\left[ n L_m^{n-m-1}(b) - b L_m^{n-m+1}(b)
  - \delta_{nm} b \right] e^{-b/2} \label{mpq} \;, \label{calmp} \eeq
which for $n\neq m$
has the same magnitude as ${\cal M}_{nm}(\bq)$, defined in \eqr{calm}.

The Hamiltonian $\cHr{eff}$, \eqr{Hph}, restricted to a single particle-hole
pair
being present, may be treated exactly as a discrete quantum state
coupled to phonons, allowing the diagrammatic computation of the
particle-hole propagator $\cal F$, \eqr{Fr}, and the spin gap.
Since momentum is conserved, consider states in which the
system has total momentum $\bQ$.
The particle's Landau level $n$ and the set of phonons present
specify a basis for the single-pair configurations.
A system which may occupy one of a discrete set of quantum states
 $|n\!\!>$,
where $n$ is a positive integer, interacting with phonons
has an isomorphic Hilbert space.
If the corresponding matrix elements are the same, the spectrum
of the discrete system agrees with the spectrum of $\cHr{eff}$
constrained to the restricted Hilbert space.
Thus, $\cHr{eff}$ with a single particle-hole pair may be transformed
into the discrete system,
\beqa \cHr{discrete} &=& \sum_{n=1}^{\infty} n\, \Psi_n^{\dagger} \Psi_n
   + \sum_{\bq} \half \hbar \omega_q
  \left( a_{\bq}^{\dagger} a_{\bq} + a_{\bq} a_{\bq}^{\dagger} \right)
\nextline
 + \sum_{n=1}^{\infty} \sum_{m=1}^{\infty} \sum_{\bq}
  \frac{ \alpha_q} {\sqrt{2\omega_q}} \, {\cal M}_{nm}'(\bq)
  \Psi_m^{\dagger} \Psi_n
 \left( a_{\bq}^{\dagger} - a_{-\bq} \right)
   \;.  \label{hdiscrete}  \eeqa
The operators $\Psi_n^{\dagger}$ and $\Psi_n$ create and destroy a
particle in the discrete state $|n\!\!>$.

We found the particle-hole propagator by solving Dyson's equation for the
 exchange of a single phonon in $\cHr{discrete}$,
as illustrated in \figr{dysonph}.
In effect, this method includes the particle exchange, hole exchange, and
particle-hole ladder Feynman diagrams.
The bare particle-hole propagator ${\cal F}^0$, defined
in \eqr{F0}, is represented by a thin dashed line,
while the phonon propagator is denoted by a thick wavy line.
The full pair propagator ${\cal F}$, represented by a thick dashed
line, may by examined in the magnetoexciton basis,
\beqa {\cal F}(\br_1 \br_4 t_1 \:|\: \br_2 \br_3 t_2) &=&
\frac{1}{2\pi} \int \sum_{n\:\bQ} \sum_{n'\:\bQ'}
 <\! n'\:\bQ' \:|\: {\cal F}(E) \:|\: n\:\bQ \!>
\nextline \mbox{\hspace {-1in}} \times
\psi_{n'\alpha'}(\br_1,\br_4) \, \psi_{n\alpha}^*(\br_2,\br_3) \,
e^{iE(t_2-t_1)} \, dE \;. \label{Fex} \eeqa
Due to the phases of ${\cal M}'_{nm}(\bq)$ in \eqr{calmp},
the integral over the direction of the phonon momentum $\bq$
 cancels in Dyson's equation whenever $n\neq n'$,
keeping the propagator diagonal.

The resulting Dyson's equation of the particle-hole Green's function
has a form similar to Eqs.~(\ref{gdiag}) through (\ref{selfeq})
for the single-particle Green's function.
The particle-hole propagator remains both diagonal
and independent of momentum in the magnetoexciton basis,
\beq <\!n' \: \bQ' \:|\: {\cal F}(E) \:|\: n \: \bQ\!> =
 \delta_{nn'} \, \delta_{\bQ\bQ'} \, {\cal F}_n(E) \;. \eeq
The magnetoexciton propagator ${\cal F}_n(E)$ for particle Landau level $n$
obeys the equations,
\beq  {\cal F}_n(E) =  \frac{1}{E - n -\Sigma_n(E) + i\eta}
\label{Fneq} \;,  \eeq
with a self-energy,
\beq  \Sigma_n(E) = i L^2 \int_{-\infty}^{\infty} \frac{d\omega}{2\pi}
 \int \frac{d\bq}{(2\pi)^2}
 \sum_{m=1}^{\infty} \, \frac{|{\cal M}_{nm}'(q)|^2 |\alpha_q|^2}
 {\omega^2 - (\omega_q - i\eta)^2} \, {\cal F}_m(E+\omega)
\;. \eeq
Performing the integral over $\omega$, using the analytic properties
of ${\cal F}$ identical to the analytic properties of $G$ leading to
\eqr{llself}, simplifies the self-energy to
\beq \Sigma_n(E) = L^2
 \int \frac{d\bq}{(2\pi)^2}
 \sum_{m=1}^{\infty} \, |{\cal M}_{nm}'(q)|^2
  \frac {|\alpha_q|^2} {2 \omega_q} \, {\cal F}_m(E-\omega_q)
 \label{selfeqF} \;.  \eeq
\eqrs{Fneq}{selfeqF} may be solved for ${\cal F}_n(E)$.

Let us compute the spin susceptibility $\chi(q,\omega)$, as defined in
\eqr{susc}, and thus find the spin gap.
Setting $\br_1$ equal to $\br_4$ and $\br_2$ equal to $\br_3$
in the particle-hole propagator ${\cal F}$, \eqr{Fex},
and Fourier transforming yields the spin susceptibility.
In terms of the magnetoexciton basis,
\beq \chi(\bQ,\omega) = \frac{2}{L^2} \sum_{n=1}^{\infty}
 \left| <\! 0 \:|\: \rho_{\bQ} \:|\: n \: \bQ \!> \right| ^2
 {\cal F}_n(\omega) \label{suscme}  \;, \eeq
where $<\! 0 \:|\: \rho_{\bQ} \:|\: n\: \bQ \!>$ is the density matrix
element between the magnetoexciton $|n\:\bQ\!>$ and the ground state,
\beq <\! 0 \:|\: \rho_{\bQ} \:|\: n \:\bQ \!>   =
 \int \! d\br_1 \, \psi_{n\alpha}^0(\br_1,\br_1) \, e^{-i\bq\cdot\br_1}
=  \frac{L}{\sqrt{2\pi n!}}
 \left( \frac{-z_{\alpha}}{\sqrt{2}} \right) ^n
  e^{-b/2} \;, \eeq
where $b=\half Q^2$ and $z_{\alpha}=iQ_x\!-\!Q_y$.
The resulting spin susceptibility is graphed in
\figr{spinsusc}
and discussed in the introduction.
This diagrammatic approach finds a spin gap $\Delta$ of $0.32 \, \hbar
\omega_c$.
The next section will apply a different computational technique to
estimate the error bars of the spin gap calculation.

\section{Second-order perturbation theory} \label{secsec}
The spin gap may also be investigated with second-order perturbation
theory.
We can thereby estimate the uncertainty arising from leaving out
the Feynman diagrams containing crossed phonon lines.
Since the coupling constant $\alpha_q$ is $\sqrt{\pi}$ in the natural units
of the problem, processes involving multiple phonons have a noticable effect,
lowering the spin gap further.
In fact, the lower spin gap energy given by second-order perturbation
theory appears to be more accurate that the value arising from summing
diagrams, as discussed at the end of Section \ref{bosesec}.
However, second-order perturbation theory cannot determine the full
spin susceptibility spectrum, but only the spin gap.

Let us compute the spin gap, the lowest energy spin-one state, using
second-order perturbation theory.
For a system with a ground state $|0\!>$ and excited states $|l\!>$
possessing energies $E_l^{(0)}$ according to a bare Hamiltonian $\cH_0$,
 an interaction $\cH_1$ shifts the ground state energy to the value
\beq E^{(2)} = E_0^{(0)}
  -\sum_l \frac{\left| <\! l \:|\: \cH_1 \:|\: 0 \!> \right|^2 }
   {E_l^{(0)} - E_0^{(0)}} \;.  \label{second} \eeq
We will find the energy shift of the spin 1
 particle-hole state $|1\:\bQ\!\!>$,
which serves as the ground state $|0\!\!>$ in \eqr{second}.
The magnetoexciton $|1 \: \bQ\!\!>$,
defined by the wavefunction in \eqr{magneto},
has a spin-up particle in Landau level~1,
a spin-down hole in Landau level 0,
and total momentum $\bQ$.
The relevant excited states $|l\!\!>$ consist of
the phonon of momentum $\bq$
and the magnetoexciton $|m \: \bQ\!-\!\bq\!>$.

Inserting these states and the interaction of $\cHr{eff}$,
\eqr{Hph},
into \eqr{second} leads to the spin gap,
\beq  \Delta^{(2)} =
\left( \epsilon_1^{(0)} - \epsilon_0^{(0)} \right) -
 L^2 \!\int\! \frac{d\bq}{(2\pi)^2} \, \frac{|\alpha_q|^2} {2 \omega_q} \,
 \sum_{m=1}^{\infty} \frac
{\left| <\!m \: \bQ\!-\!\bq \:|\: \jt_{\bq} \:|\: 1 \: \bQ \!> \right|^2 }
   {\epsilon_m^{(0)} \!-\! \epsilon_1^{(0)} + \omega_q }
\label{phen}  \;. \eeq
Replacing the current matrix element in \eqr{phen}
with ${\cal M}_{m1}'(\bq)$, as defined in \eqr{calmp},
and using the values in \eqr{en0} for $\epsilon_n^{(0)}$
 yields the lowest energy of a spin-one excitation,
\beq  \Delta^{(2)} =
1- L^2 \!\int\! \frac{d\bq}{(2\pi)^2}  \, \frac{|\alpha_q|^2} {2 \omega_q}
 \sum_{m=1}^{\infty} \frac {|{\cal M}_{m1}'(\bq)|^2} { m - 1 + \omega_q }
= 1 - 0.83 \;. \label{egap2} \eeq
Thus, at the level of second-order perturbation theory,
 the phonon interaction reduces the spin gap
to $0.17 \,\hbar \omega_c$, less than 20\% of its mean-field value,
compared to a spin gap of $0.32 \,\hbar \omega_c$ obtained by
diagrammatic methods in Section \ref{parthole}.

Altering the sound speed $v_s$ in the phonon dispersion, \eqr{omegaqf},
changes the energy gap significantly, increasing the error bars on
the spin gap calculation.
For instance, a 10\% decrease in the sound speed to $v_s=0.9$
reduces the second-order perturbation theory energy gap
calculated in \eqr{egap2} to
\beq  \Delta^{(2)\ v_s=0.9} = 1 - 0.96 \;. \eeq
Since such a sound speed is plausible~\cite{hannacorr},
we cannot set a lower
bound on the size of the spin gap, or even prove that a spin gap
exists.
Still, the spin gap definitely lies in the range
\beq \Delta = (0.2 \pm 0.2) \, \hbar \omega_c \;, \eeq
much less than the mean-field value of $\hbar \omega_c$.

\section{Boson test case}   \label{bosesec}

We will compute the Green's function of the non-interacting spinless
bose gas in order to test the computational procedure used to find
the spin gap of the spin $\half$ fractional-statistics gas.
The same gauge transformation changing particles with
fractional statistics into fermions obeying $\cH_{\rm anyon}$, \eqr{anyonH},
will also transform a gas of non-interacting bosons,
\beq  \cH_{\rm bose} = \sum_{j=1}^N \frac{|{\bf P}_j|^2}{2m}
\;, \label{boseH} \eeq
into $\cH_{\rm anyon}$ with a statistical factor $\nu\!=\!0$,
and with the cutoff parameter $\alpha\!\rightarrow\!\infty$.
Since the bose energy eigenstates are known exactly,
the bose Green's function may also be computed exactly.
Agreement between the exact and approximate
bose Green's functions demonstrates the validity of the semion
calculation, because bose statistics involve a larger deviation
from non-interacting fermions than do fractional statistics.

We shall compute the part of the zero-distance bose Green's function,
\beq  G(E) = -i \int_{-\infty}^{\infty} e^{iEt} \,
  < \Phi \:|\: T \{ \Psi(0,t) \, \Psi^{\dagger}(0,0) \} \:|\: \Phi >
dt \;, \label{zdgreen}  \eeq
arising from the occupied states,
where $|\Phi\!\!>$ is the ground state.
In the fermionic representation,
$\Psi^{\dagger}(\br)$
and $\Psi(\br)$ are local fermionic operators
creating or destroying a fermion at position $\br$.
In the bosonic representation, however, those operators
also multiply the wavefunction by a phase depending upon the positions
of all the other particles, effectively creating or destroying a vortex.
Thus, the Green's function defined in \eqr{zdgreen} differs substantially
from the traditional bose propagator.
In terms of the excited states $|l\!>$ with energies $\epsilon_l$,
the Green's function has the value,
\beq G(E) = \sum_l \left\{
 \frac{\left| <\! l \:|\: \Psi^{\dagger}(0) \:|\: \Phi \!> \right| ^2 }
    {E - \epsilon_l + i\eta}  +
 \frac{\left| <\! l \:|\: \Psi(0) \:|\: \Phi \!> \right| ^2 }
    {E + \epsilon_l - i\eta}
\right\}  \;.  \eeq
The imaginary part of the Green's function corresponding to
 the occupied states is then,
\beq \frac{1}{\pi} \, {\rm Im} \, G(E) = \sum_l
 \left| <\! l \:|\: \Psi(0) \:|\: \Phi \!> \right| ^2
  \delta(E + \epsilon_l)  \label{imgeb}  \;, \eeq
valid for $E<0$, which are the only energies we shall consider.

The Green's function calculated with the approximation techniques used
to find the spin gap of the fractional-statistics gas agrees fairly
well with the exact boson spectrum, as displayed in \figr{gbose}.
We are using units in which
$a_0$, \eqr{a0}, and $\omega_c$, \eqr{omegac}, defined with $\nu=0$,
are both set equal to one.
The exact bose Green's function, the solid line in \figr{gbose},
was found by substituting the exact bose wavefunctions and energies
into \eqr{imgeb}.
The approximate bose Green's function, the dashed line in \figr{gbose},
was found by first determining the bose effective Hamiltonian in the form
of $\cHr{eff}$, \eqr{Hph}.
We then considered, in addition to the single-phonon exchange processes
considered in Section \ref{singpart},
processes involving arbitrary numbers of crossed phonon interaction lines,
such as the diagram in \figr{crossedlines}.
The effect of such crossed phonon graphs in the fractional-statistics
gas is discussed at the end of this section.
The exact and approximate Green's functions possess a similar overall
structure.  In addition,
the highest energy hole states in the two calculations differ in
energy by only $0.06 \, \hbar\omega_c$,
demonstrating the reliability
of the calculational technique used to compute the spin gap
 of the fractional-statistics gas.

Let us compute the bose Green's function exactly.
The system consists of $N$ bosons confined to a region of size $L^2$,
with an average density,
\beq  \bar{\rho} = \frac{N}{L^2} = \frac{1}{2\pi} \;. \eeq
The gauge transformation
from $\cHr{bose}$, \eqr{boseH}, to $\cHr{anyon}$, \eqr{anyonH},
multiplies the bose wavefunction by a phase depending upon the
direction of the separation vector between each pair of particles,
but leaves the excitation energies unchanged.
In the fermionic representation, the $N$ particle ground state
$|\Phi\!\!>$ has the wavefunction,
\beq |\Phi\!> =  \frac{1}{L^N}
\prod_{i<j}^N \left(  \frac{z_i \!-\! z_j} {|z_i \!-\! z_j|}  \right)
  \label{bgs} \;, \eeq
where $z=x+iy$ expresses a position as a complex number.
The $N\!-\!1$ particle excited states
$|{\bf k}_1,\ldots,{\bf k}_{N\!-\!1} \!>$ consist of plane waves,
with energies
\beq  \epsilon_{{\bf k}_1,\ldots,{\bf k}_{N\!-\!1}} =
  \sum_{j=1}^{N-1} \half k_j^2   \label{exactbe} \eeq
and wavefunctions
\beq  | {\bf k}_1,\ldots,{\bf k}_{N\!-\!1}\!> =
\prod_{i<j}^{N-1} \left( \frac{z_i \!-\! z_j} {|z_i \!-\! z_j|}
\right) {\cal N} \sum_{\tau}^{(N-1)!}
e^{i \left[ {\bf k}_1 \cdot \br_{\tau(1)} + \cdots +
  {\bf k}_N \cdot \br_{\tau(N\!-\!1)} \right]}
    \label{bes}  \eeq
summed over permutations $\tau$.
The normalization ${\cal N}$,
\beq {\cal N} = \frac{1}{L^{N-1} \sqrt{(N\!-\!1)! \, n_1! \cdots n_M!} }
  \;, \eeq
depends upon the momenta ${\bf k}_j$ with an occupancy $n_j$
greater than one.

The bose Green's function may then be calculated from the energy eigenstates.
The matrix element in \eqr{imgeb}
coupling the ground state $|\Phi\!\!>$, \eqr{bgs}, to the
excited state $| {\bf k}_1,\ldots,{\bf k}_{N\!-\!1}\!\!>$, \eqr{bes},
 has the magnitude,
\beq \left| <\! {\bf k}_1,\ldots,{\bf k}_{N\!-\!1}
  \:|\: \Psi(0) \:|\: \Phi \!> \right| ^2
 =  \frac{{\cal N}^2}{L^{2N}}
 \prod_{j=1}^{N\!-\!1}
\left(  \frac {2\pi} {k_j^2}  \right)^2
\;.  \eeq
Substituting these matrix elements and the excitation energies of
\eqr{exactbe} into \eqr{imgeb} leads to,
\beqa \frac{1}{\pi} \, {\rm Im} \, G(E) & = &
 \frac{1}{L^{2N}} \left( {2\pi} \right) ^ {2(N-1)} \!\!
 \sum _{ \left\{ {\bf k}_1, \ldots, {\bf k}_{N\!-\!1} \right\} }
 \frac{{\cal N}^2} { k_1^4 \cdots k_{N\!-\!1}^4 } \,
 \delta\! \left( E + \sum_{j=1}^{N\!-\!1} \half |{\bf k}_j|^2 \right)
\nexteq
 \frac{1}{L^{2N}} \int
 \frac{1} { k_1^4 \cdots k_{N\!-\!1}^4 } \,
 \delta\! \left( E + \sum_{j=1}^{N\!-\!1} \half |{\bf k}_j|^2 \right) \,
 d{\bf k}_1 \cdots d{\bf k}_{N\!-\!1}
\;.  \eeqa
Changing the sum into an integral generates a combinatoric
factor cancelling out a similar term in the normalization ${\cal N}$.
To avoid an infared divergence, we will cut off the momentum
at a value $k_0$, requiring a normalization of ${\rm Im} \, G(E)$
consistent with the density $\bar{\rho}$,
\beq  \bar{\rho} =
 \int_{-\infty}^{0} \frac{1}{\pi} \, {\rm Im} \, G(E) \, dE
= \frac{1}{L^{2N}} \int_{k_j>k_0}
 \frac{1} {k_1^4 \cdots k_{N\!-\!1}^4} \,
 d{\bf k}_1 \cdots d{\bf k}_{N\!-\!1}  \;,  \eeq
so that
\beq  k_0 = \frac{\sqrt{\pi}}{L}  \;.  \eeq
With the cutoff,
one can take the Fourier transform of ${\rm Im} \, G(E)$,
\beqa \int_{-\infty}^{0}
   e^{iEt} \, \frac{1}{\pi} \, {\rm Im} \, G(E) \, dE & = &
 \frac{1}{L^{2N}} \left( \int_{k>k_0} \frac{1}{k^4}
  e^{i\half t k^2} d{\bf k} \right) ^{N-1}
\nexteq
 \bar{\rho} \left\{ \left. \left(
\int_{\half k_0^2}^{\infty}
\frac{1}{\epsilon^2} \, e^{it\epsilon} \, d\epsilon
\right) \right/  \left(
 \int_{\half k_0^2}^{\infty} \frac{1}{\epsilon^2} \, d\epsilon \right)
    \right\} ^{N-1}
\nexteq
 \bar{\rho} \left\{ \left[ \frac{1}{\half k_0^2} - it(\gamma - 1)
  -it \ln \left(-it \half k_0^2 \right) \right]
   \half k_0^2 \right\} ^ {N-1}
\;. \eeqa
Transforming back to $E$ yields the exact hole Green's function,
\beq \frac{1}{\pi} \, {\rm Im} \, G^{\rm exact}(E) =
 \frac{\bar{\rho}}{2\pi} \int_{-\infty}^{\infty}
   e^{(i/4) t[\ln|t| - 1]} e^{-(\pi/8) |t|}
  \, e^{-i (E-\epsilon_0^{\rm exact}) t} \, dt \;, \label{gexact} \eeq
where
\beq  \epsilon_0^{\rm exact} =
  -\half \ln (R) - \frac{1}{4} \left[ \ln (2) - \gamma \right] \;, \eeq
with $L^2 \!=\! \pi R^2$ and $\gamma$ denoting Euler's constant.
The exact Green's function is graphed as the solid line in \figr{gbose}.

Let us now compute the hole propagator of the bosons
with a technique
similar to that used in Section \ref{singpart} for semions,
in order to test our approximation method in comparison with the
exact result $G^{\rm exact}(E)$, \eqr{gexact}.
Some changes are necessary in the bose case.  In particular,
the collective mode of the bose gas has a quadratic dispersion,
\beq  \omega_q = \half q^2 \;. \label{omegaqb} \eeq
Even though the RPA calculation of Section \ref{effintsec}
for a bose gas obtains a linear dispersion,
we will use the quadratic dispersion,
which would arise if all Feynman diagrams were included.
However, the effective interaction ${\cal V}$ between two bosons
in the fermionic representation will still be approximated by \eqr{VVDV},
which was used in Section \ref{effintsec} to compute the effective
interaction of the fractional-statistics gas.
In that formula,
the bare potential ${\cal V}^0$ for bosons is twice as large
as the semion case value, \eqr{V0}.
We will again approximate the effective interaction
as the bare potential ${\cal V}^0$
added to a sharp mode at each momentum $\bq$ in the
transverse current-transverse current channel, as in \eqr{veff}.
The polarization bubble ${\cal D}$ of the bose gas,
defined in \eqr{corrfun},
satisfies the sum-rule argument in Section \ref{effintsec},
resulting in a coupling strength of
\beq |\alpha_q|^2 = \frac{2\pi}{L^2} \;.  \label{alphaqb} \eeq
The static interaction ${\cal V}^0$ moves each Landau level $n$
from the bare energy $\epsilon_n^{(0)}$, \eqr{en0},
 to its Hartree-Fock energy $\epsilon_n^{\rm HF}$,
which for Landau level 0 has the value
\beq \epsilon_0^{\rm HF} =
  -\half \ln \,(R) + \frac{1}{4} [\ln\, (2) - \gamma]
\label{e0hf} \eeq
for a sample of size $L^2 \!=\! \pi R^2$.
In summary, the resulting system is described by the effective Hamiltonian
$\cH_{\rm eff}$, \eqr{Hph}, with the phonon energy $\omega_q$
given by \eqr{omegaqb}
and the coupling $\alpha_q$ given by \eqr{alphaqb}.
The energies corresponding to Landau level 0 should be shifted by
the amount $\epsilon_0^{\rm HF} \!-\! \epsilon_0^{(0)}$.

Because the phonons have a quadratic dispersion,
processes with crossed phonon lines, such as those in \figr{crossedlines},
make a significant contribution to the bose spectrum.
The solution to Dyson's equation for the exchange of a single phonon,
 \eqrs{Gneq}{llself0} does not determine the propagator accurately.
Instead,
we will sum the Feynman diagrams with any number of crossed phonon
lines.
In the last paragraph of this section, we will discuss how such
multi-phonon graphs affect the fractional-statistics gas.
Since level 0 is the only occupied Landau level,
we shall only consider graphs where the hole lies in Landau level 0
in the initial, final, and all intermediate states.

To find the bose Green's function,
a single Landau level interacting with phonons will be treated as a
single quantum state interacting with phonons.
The effective Hamiltonian $\cHr{eff}$, \eqr{Hph}, representing the bose
gas in the fermionic representation, \eqr{anyonH}, will in turn be
transformed into the discrete state Hamiltonian
\beq \cH_{\rm single} =
 \sum_{\bq} \half \omega_q
\left( a_{\bq} a_{\bq}^{\dagger} + a_{\bq}^{\dagger} a_{\bq} \right)
+ \sum_{\bq} {\lambda}_q \Psi^{\dagger} \Psi
  \left( a_{\bq} + a_{-\bq}^{\dagger} \right)
\label{Hsingle}  \;,  \eeq
similar to the Hamiltonian $\cHr{discrete}$, \eqr{hdiscrete}, utilized
to find the particle-hole propagator of the fractional-statistics gas
in Section \ref{parthole}.
The operators $\Psi^{\dagger}$ and $\Psi$ create and destroy a fermion
which may occupy a single quantum state,
$a_{\bq}^{\dagger}$ and $a_{\bq}$ create and destroy a phonon,
and the fermion-phonon coupling ${\lambda}_q$ has the value
\beq |{\lambda}_q|^2
= \frac{|\alpha_q|^2}{2\omega_q} |{\cal M}_{00}(q)|^2
 = \frac{\pi}{2L^2} e^{-\half q^2}  \;,  \eeq
where $\omega_q$ and $\alpha_q$ have the bosonic values of
\eqrs{omegaqb}{alphaqb} and $\cal M$ is defined in \eqr{calm}.
The Green's function of $\cHr{single}$,
\beqa G^{\rm single}(t_1,t_2) & = &
  -i < \Phi \:|\: T \{ \Psi(t_1) \: \Psi^{\dagger}(t_2) \}  \:|\: \Phi >
\nexteq
    \int_{-\infty}^{\infty} G(E) \, e^{iE(t_2-t_1)} \, dE
    \;, \label{greens}  \eeqa
where $|\Phi\!\!>$ is the vacuum state,
may be converted into
an approximate zero-distance bose Green's function
defined in \eqr{zdgreen},
\beq {\rm Im} \, G^{\rm approx}(E) =
 -{\bar{\rho}} \,
{\rm Im} \, G^{\rm single} (\epsilon_0^{\rm HF} - E)
\label{gconv} \;. \eeq
The differences between a discrete quantum state
and a hole in any orbital of
a filled Landau level account for the changes in sign and normalization.
For processes involving the exchange of a single phonon,
the self-energy expression for $\cHr{single}$ agrees with
the single-hole self-energy formula \eqr{selfeq}
with $n$ and $m$ both set to Landau level 0.
For the process involving two crossed phonon lines, as in
\figr{crossedlines},
the spatial integral due to the Hamiltonian $\cHr{eff}$, \eqr{Hph},
 reduces to the sum of
 the matrix elements of transverse-current operator $\jt_{\bq}$, \eqr{jqt},
between the Landau level orbitals in \eqr{llstates},
\beqa & & \mbox{\hspace{-1in}} \sum_{p p' p''}
<\! 0 \: k' \:|\: \jt_{-\bq'} \:|\: 0 \: p \!>
<\! 0 \: p \:|\: \jt_{-\bq} \:|\: 0 \: p' \!>
<\! 0 \: p' \:|\: \jt_{\bq'} \:|\: 0 \: p'' \!>
<\! 0 \: p'' \:|\: \jt_{\bq} \:|\: 0 \: k \!>
\nexteq
 \delta_{kk'} e^{i(\bq' \times \bq) \cdot \zhat}
|{\cal M}_{00}(q)|^2 |{\cal M}_{00}(q')|^2
  \;, \eeqa
Since small momentum phonons have the dominant effect,
the phase $e^{i (\bq' \times \bq) \cdot \zhat}$ may be neglected.
Then the two-phonon Feynman diagrams for both $\cH_{\rm eff}$
and $\cH_{\rm single}$ have the same self-energy expression,
\beqa  \Sigma(E)_{\rm (2\ phonon)}  & = &
 - \int_{-\infty}^{\infty} \frac{d\omega}{2\pi}
 \int \frac{d\bq}{(2\pi)^2}
  \int_{-\infty}^{\infty} \frac{d\omega'}{2\pi}
 \int \frac{d\bq'}{(2\pi)^2}
\nextline \times
  |{\cal M}_{00}(q)|^2 |{\cal M}_{00}(q')|^2
  \frac {|\alpha_p|^2} {(\omega + i\eta)^2 - \omega_q^2}
  \frac {|\alpha_q|^2} {(\omega' + i\eta)^2 - \omega_{q'}^2}
\nextline \times
  G_0(E+\omega) \, G_0(E+\omega+\omega') \, G_0(E+\omega')
 \label{selfeq2} \;.  \eeqa
This agreement may be generalized to Feynman diagrams with any
number of crossed phonon lines.  Thus, we may study $\cHr{single}$
to find the hole Green's function of the bose gas in the fermionic
representation.

The Hamiltonian $\cH_{\rm single}$, \eqr{Hsingle},
whose Green's function is readily converted into the zero-distance
Bose green's function,
may be solved exactly by transforming
the phonons, treated as harmonic oscillators, into the conjugate
operators $x_{\bq}$ and $P_{\bq}$ at each momentum $\bq$.
The phonon creation and annihilation operators $a_{\bq}^{\dagger}$
and $a_{\bq}$ are the standard linear combinations of
$x_{\bq}$ and $P_{\bq}$,
\beq a_{\bq} = \frac{1}{\sqrt{2\omega_q}}
 \left( \sqrt{K_q}\,x_{\bq} + \frac{iP_{\bq}}{\sqrt{m_q}} \right) \;, \eeq
and
\beq a_{\bq}^{\dagger} = \frac{1}{\sqrt{2\omega_q}}
 \left( \sqrt{K_q}\,x_{\bq} - \frac{iP_{\bq}}{\sqrt{m_q}} \right) \;, \eeq
where $K_q$ and $m_q$ satisfy
\beq  \omega_q = \sqrt{\frac{K_q}{m_q}}  \;.  \eeq
The Hamiltonian $\cH_{\rm single}$ may then be reexpressed as,
\beqa \cH_{\rm single} & = &
 \sum_{\bq} \left[ \half K_q x_{\bq}^2 + \frac{P_{\bq}^2}{2m_q}  \right]
 + \sum_{\bq} \lambda_q \sqrt{\frac{2K_q}{\omega_q}}
    x_{\bq} \Psi^{\dagger} \Psi
\nexteq
 \sum_{\bq} \left[ \half K_q \left( x_{\bq} +
   \lambda_q \sqrt{\frac{2}{K_q\omega_q}} \Psi^{\dagger} \Psi \right) ^2
 \right]
  + \Delta E \, \Psi^{\dagger} \Psi
\;. \eeqa
When the discrete quantum state is occupied,
its interaction shifts the harmonic oscillators displacements and
lowers the total energy by
\beq \Delta E = -\sum_{\bq} \frac {\lambda_q^2} {\omega_q}
\label{deltae} \;. \eeq
The vacuum state is then
\beq |\Phi\!> = \prod_{\bq} e^{-\half \left( x_{\bq}/d_q \right) ^2}
\;, \eeq
with a length scale
\beq d_q = (K_q m_q)^{-1/4} \;. \eeq
A relevant excited state contains the discrete state and $n_{\bq}$ phonons
of each momentum $\bq$, with the shifted harmonic oscillator wavefunction
\beq |n_{\bq}\!> = \Psi^{\dagger}
 \prod_{\bq} \left[ {2^{n_{\bq}} n_{\bq}! \sqrt{\pi} d_q} \right] ^{-1/2}
   H_{n_{\bq}} \! \left( \frac{1}{d_q}
 \left[{x_{\bq}} + \lambda_q \sqrt{\frac{2}{K_q\omega_q}} \, \right]
\right)
e^{-\half \left( x_{\bq}/d_q \right) ^2}  , \label{wnq} \eeq
where the $H_n$ are Hermite polynomials,
and an energy,
\beq \epsilon_{ n_{\bq} } =
\Delta E + \sum_{\bq} n_{\bq} \, \omega_q
 \;. \label{enq} \eeq

The excited states $|n_{\bq}\!\!>$, \eqr{wnq},
and energies $\epsilon_{n_{\bq}}$, \eqr{enq},
determine the imaginary part of the Green's function, defined
in \eqr{greens},
\beq -\frac{1}{\pi} \, {\rm Im} \, G^{\rm single}(E) =
\sum_{\{n_{\bq}\}}
  \left| <\! n_{\bq} \:|\: \Psi^{\dagger} \:|\: \Phi \!> \right| ^2
\, \delta\! \left( E - \epsilon_{n_{\bq}} \right) \;. \label{imgea} \eeq
The overlap between $\Psi^{\dagger}|\Phi\!\!>$ and
the excited state $|n_{\bq}\!\!>$ has a magnitude,
\beq \left| <\!n_{\bq} \:|\: \Psi^{\dagger} \:|\: \Phi\!> \right| ^2 =
 \prod_{\bq} \frac{c_q^{n_{\bq}}} {n_{\bq}!} e^{-c_q}
\label{overlap} \;, \eeq
where $c_q$ stands for,
\beq c_q = \left( \frac{\lambda_q}{\omega_q} \right) ^2
 = \frac{2\pi}{L^2q^4} e^{-\half q^2} \;. \label{cq} \eeq
The energy shift $\Delta E$ in \eqr{deltae} diverges due to phonons
of small momentum.  However, the state $\Psi^{\dagger}|\Phi\!\!>$
has an expected energy of
\beq  <\!{\cH_{\rm single}}\!> =
 \Delta E + \sum_{\bq} <\!{n_{\bq}}\!> \omega_q =
0  \;, \eeq
where $<\!{\cal O}\!>$ is shorthand for
$<\!\!\Phi \,|\, \Psi \, {\cal O} \, \Psi^{\dagger} \,|\, \Phi\!\!>$.
Since on average the excited phonons cancel the energy shift,
the energy of a state depends upon the fluctuations,
\beq  \epsilon_{ n_{\bq} } =
 \sum_{\bq} \, (n_{\bq} - <\!{n_{\bq}}\!> ) \, \omega_q
\label{enfluct}  \;. \eeq
Substituting \eqr{overlap} and \eqr{enfluct} into \eqr{imgea}
and taking the Fourier transform leads to
\beqa
 \int_{-\infty}^{\infty} e^{-iEt}  \left( -\frac{1}{\pi}
\, {\rm Im} \, G^{\rm single}(E)  \right) &=&
\sum_{\{n_{\bq}\}} \, \prod_{\bq} \, \frac{c_q^{n_{\bq}}} {n_{\bq}!}
\, e^{-c_q} \,  e^ { -i (n_{\bq} - c_q) \, \omega_q t }
\nexteq
  \exp \left\{ \sum_{\bq} c_q
\left( e^{-it\omega_q} \!-\! 1 \!+\! it\omega_q \right) \right\}
\nexteq
e^{ (1/4) \, [ (1+it) \ln\, (1+it) - it] }
 \;, \label{gsingft}  \eeqa
where the imaginary part of the logarithm lies in the range,
\beq -\frac{\pi}{2} < {\rm Im} \left[ \ln\,(1+it) \right]
  < \frac{\pi}{2}  \;.  \eeq
After Fourier transforming back to energy dependence
 and shifting the $t$ contour
 in the complex plane, the Green's function of $\cH_{\rm single}$ becomes,
\beq -\frac{1}{\pi} \, {\rm Im} \, G^{\rm single}(E) =
 \frac{1}{2\pi} e^{1/4 - E }
\int_{-\infty}^{\infty}  e^{(i/4)t [\ln|t| - 1]} \, e^{-(\pi/8) |t|} \,
e^{i E t} \, dt \label{gsingle}  \;. \eeq
The approximate zero-distance Green's function for the hole states,
after applying \eqr{gconv},
\beq \frac{1}{\pi} \, {\rm Im} \, G^{\rm approx}(E)   =
 \frac{\bar{\rho}} {2\pi} e^{1/4 + (E - \epsilon_0^{\rm HF}) }
\int_{-\infty}^{\infty}  e^{(i/4) t[\ln|t| - 1]} \, e^{-(\pi/8) |t|} \,
e^{-i (E - \epsilon_0^{\rm HF}) t}  \, dt \label{gapprox}  \;, \eeq
is plotted in \figr{gbose} as a dashed line,
and displays good agreement with the exact spectrum
 $G^{\rm exact}(E)$, \eqr{gexact}.

The calculation of the spin gap in Section \ref{parthole} left out
the Feynman diagrams with crossed phonon lines.
The method used in this section to deal with those graphs fails
when Landau level excitations are involved.
Although the higher Landau states affect the spin susceptibility,
interactions involving only Landau levels 0 and 1 cause a majority
of the shift of the spin gap.
If the Landau levels $n\geq 2$ are neglected, the preceding analysis
of the fractional-statistics lowers the gap by the amount $\Delta E$
in \eqr{deltae}.
Then, since $\Delta E$ is simply the
second-order perturbation theory result
found in Section \ref{secsec}, second-order perturbation theory
 gives the same spin gap as does including all the Feynman diagrams,
for the Landau levels causing most of the gap's shift.
Thus, the spin gap found by second-order perturbation theory is
more accurate than the diagrammatic calculation in Section \ref{parthole}.

\acknowledgments
The authors are grateful to A.~Fetter and S.~Girvin for helpful discussions.
This work was supported primarily by the National Science Foundation
under Grant No.\ DMR-88-16217.  Additional support was provied by the NSF
MRL program through the Center for Materials Research at Stanford University.
JLL wishes to thank the John and Fannie Hertz Foundation for
fellowship support.

\begin{figure}
\caption{Illustration of the mean-field density of states of the spin $\half$
fractional-statistics gas.
Landau level 0 is occupied with particles of both spins.}
\label{landaul}
\end{figure}

\begin{figure}
\caption{Imaginary part of the spin susceptibility $\chi(q,\omega)$,
defined in {\protect\eqr{susc}},
at three momenta:  (a) $Q=0.2/a_0$, (b) $Q=0.4/a_0$, and (c) $Q=0.6/a_0$,
as calculated using {\protect\eqr{suscme}}.
The vertical lines (d) display the delta-function mean-field
susceptibility at $Q=0.6/a_0$.}
\label{spinsusc}
\end{figure}

\begin{figure}
\caption{Illustration of decay processes causing broadening of the
single-particle spectrum:
(a) interlevel transitions and (b) intralevel transitions.
Each decay process emits a phonon.}
\label{decayproc}
\end{figure}

\begin{figure}
\caption{Diagrammatic representation of the effective interaction:
(a) exact expression in terms of the proper polarization bubble ${\cal D}^P$,
as in \protect\eqr{VV2}, and
(b) RPA approximate expression in terms of the bare polarization bubble
${\cal D}^0$, as in \protect\eqr{vrpa}. }
\label{effint}
\end{figure}

\begin{figure}
\caption{Imaginary part of ${\cal V}_{JJ}$, the
 transverse current-transverse current component
 of the RPA effective interaction given in {\protect\eqr{vrpaf}},
at three momenta:  (a) $q=0.5/a_0$, (b) $q=1/a_0$, and (c) $q=4/a_0$.
The solid lines are broadened with a small $\eta$ for visibility, while
the dashed lines of (b) and (c) are broadened with a larger $\eta$ to resemble
the full effective interaction.
The collective mode energy $\omega_q$ is defined in {\protect\eqr{omegaave}}
to be the average value of $\omega$,
and the coupling $|\alpha_q|^2$ defined in {\protect\eqr{alphave}}
 is proportional to the integrated interaction strength
at a given $q$.  }
\label{imvrpa}
\end{figure}

\begin{figure}
\caption{Collective mode dispersion $\omega_q$.
The solid line is the averaged RPA interaction defined by
\protect\eqr{omegaave},
and the dashed line is the approximation formula
given in {\protect\eqr{omegaqf}}.}
\label{omegaqfig}
\end{figure}

\begin{figure}
\caption{Typical Feynman diagram involving the static interaction,
 the ${\cal V}^0$ of {\protect\eqr{veff}},
 which as discussed at
the end of Section {\protect\ref{effintsec}} does not affect the spin gap. }
\label{kohn}
\end{figure}

\begin{figure}
\caption{Illustration of Dyson's equation due to phonon exchange
for the single-particle Green's function $G$,
as expressed in Eqs.~({\protect\ref{gllbasis}}) to ({\protect\ref{selfeq}}).
The labels on the particle lines refer to the Landau level orbitals given in
\protect\eqr{llstates}.}
\label{dysons}
\end{figure}

\begin{figure}
\caption{Contour used to evaluate the $\omega$ integral in the
self-energy expression $\Sigma_n(E)$, {\protect\eqr{selfeq}},
for Landau levels $n\geq 1$ and $m\geq 1$.}
\label{contour}
\end{figure}

\begin{figure}
\caption{Density of states $D_n(E)$, as defined in {\protect\eqr{dos}},
for Landau levels 0 to 3.
The bare Landau level energies $\epsilon_n^{(0)}$
 given in {\protect\eqr{en0}} are indicated on the top of the graph.
Neglecting the particle-hole interactions,
the spin gap is the difference between the maximum hole energy $E_{\rm max, h}$
and the minimum particle energy $E_{\rm min, p}$.
As discussed following {\protect\eqr{dos}}, the calculation
leaves out the divergent gap separating the occupied and unoccupied
states.}
\label{imge}
\end{figure}

\begin{figure}
\caption{Diagrammatic sum used to compute the spin one particle-hole
propagator {\protect{$\cal F$}}, defined in {\protect\eqr{Fr}}.
The particle-hole pair representation is transformed into a discrete system
obeying $\cHr{discrete}$, {\protect\eqr{hdiscrete}}.
The indices on the dashed lines
refer to either the particle Landau level in the
magnetoexciton wavefunctions given in
{\protect\eqr{magneto}} or the states of $\cHr{discrete}$. }
\label{dysonph}
\end{figure}

\begin{figure}
\caption{Green's function $G$, defined in {\protect\eqr{zdgreen}},
for the holes states of the non-interacting
bose gas in the fermionic representation.
The solid line denotes the
exact spectrum $G^{\rm exact}(E)$, given by {\protect\eqr{gexact}},
and the dashed line denotes the approximate spectrum $G^{\rm approx}(E)$,
given by {\protect\eqr{gapprox}}.
The energy $E=0$ on the graph corresponds to $\epsilon_0^{\rm HF}$,
the Hartree-Fock energy of Landau Level 0 given in
{\protect\eqr{e0hf}}. }
\label{gbose}
\end{figure}

\begin{figure}
\caption{Illustration of a process involving two crossed phonon interaction
lines included in the bose gas calculation, as
expressed in {\protect\eqr{selfeq2}}.
The hole propagators are labeled by the Landau level orbitals
defined in {\protect\eqr{llstates}}.}
\label{crossedlines}
\end{figure}


\begin{references}

\bibitem{overview} R.\ B.\ Laughlin, Science {\bf 242}, 4878 (1988);
 Y.-H. Chen, F.\ Wilczek, E.\ Witten, and B.~I.\ Halperin,
  Int.\ J.\ Mod.\ Phys.\ B {\bf 3}, 1001 (1989);
 R.\ B.\ Laughlin, to be published in
 {\it Modern Perspectives in Many-Body Physics}, edited by M.\ P.\ Das
  and J.\ Mahanty (World Scientific Publ., Singapore).

\bibitem{tikofsky}  A.\ M.\ Tikofsky, R.\ B.\ Laughlin, and Z. Zou,
 Phys.\ Rev.\ Lett. {\bf 69}, 3670 (1993).

\bibitem{experiment}  {J. Rossat-Mignod {\it et al.}, Physica~B {\bf 12},
 109 (1993);
 Sato {\it et al.}, J.\ Phys.\ Soc.\ Jap. {\bf 62} 263 (1993). }

\bibitem{girvin} S.\ M.\ Girvin, A.\ H.\ MacDonald, M.\ P.\ A.\ Fisher,
 S.-J.\ Rey, and J.\ P.\ Sethna, Phys.\ Rev.\ Lett.
 {\bf 65}, 1671 (1990).

\bibitem{hannahf} C.\ B.\ Hanna, R.\ B.\ Laughlin, and A.\ L.\ Fetter,
 Phys.\ Rev.\ B {\bf 40}, 8745 (1989).

\bibitem{beran}  P.\ B\'{e}ran and R.\ B.\ Laughlin, Phys.\ Rev.\ B {\bf 48},
 10382 (1993).

\bibitem{dai} Q.\ Dai, J.\ L.\ Levy, A.\ L.\ Fetter, C.\ B.\ Hanna, and
  R.\ B.\ Laughlin, Phys.\ Rev.\ B {\bf 46}, 5642 (1992).

\bibitem{fetterrpa} A.\ L.\ Fetter, C.\ B.\ Hanna, and R.\ B.\ Laughlin,
 Phys.\ Rev.\ B, {\bf 39}, 9679 (1989).

\bibitem{hannacorr} C.\ B.\ Hanna and A.\ L.\ Fetter, Phys.\ Rev.\ B {\bf 47},
 3280 (1993).

\bibitem{pines} D.\ Pines and P.\ Nozieres, {\it The Theory of Quantum Liquids}
 (Benjamin, New York, 1966).

\bibitem{kohn} W.\ Kohn, Phys.\ Rev.\ {\bf 123}, 1242 (1961).

\bibitem{tammdancoff} I.\ Tamm, J.\ Phys.\ (U.S.S.R.) {\bf 9}, 449 (1945);
 S.\ M.\ Dancoff, Phys. Rev. {\bf 78}, 382 (1950);
 H.\ A.\ Bethe and F. De Hoffmann, {\it Mesons and Fields}
  (Row, Peterson, and Company, Evanston, Illinois, 1955);
 A.\ L.\ Fetter and J.\ D.\ Walecka, {\it Quantum Theory of Many Body
Systems} (McGraw-Hill, New York, 1971).

\bibitem{hannaph} C.\ B.\ Hanna, R.\ B.\ Laughlin, and A.\ L.\ Fetter,
 Phys.\ Rev.\ B {\bf 43}, 809 (1991).

\end{references}
\end{document}